\documentclass[12pt]{article}
\usepackage{bm}% bold math
\title{Can effective muon $g-2$ depend on the gravitational potential?}
\author{Hrvoje Nikoli\'c \\
Theoretical Physics Division, Rudjer Bo\v{s}kovi\'{c} Institute, \\
P.O.B. 180, HR-10002 Zagreb, Croatia \\
{\normalsize e-mail: hnikolic@irb.hr} \\
\makebox[1in]{} \\
}
\date{\today}
\begin{document}
\maketitle
\begin{abstract}
Contrary to the claim in a series of recent papers, we show that it cannot. 
A source of the error in those papers is misinterpretation of coordinate time as a physical time.  
\end{abstract}
\vspace*{0.5cm}
%PACS Numbers: 04.20.-q, 04.20.Cv
%04.20.-q 	Classical general relativity 
%04.20.Cv 	Fundamental problems and general formalism

\noindent
In a recent series of papers \cite{p1,p2,p3} the authors argued that gravitational potential of the Earth
has influence on the effective anomalous magnetic moment $g-2$ of muons. Visser objected \cite{visser} that it cannot be right
because the dependence on the potential would contradict the equivalence principle of general relativity.
We agree with Visser and identify an exact place of error in the analysis presented in \cite{p1,p2,p3}.

In \cite{p1}, the authors consider a charged particle in magnetic field ${\bm B}$ moving in the horizontal plane. 
They find that such a motion is described by the equation 
\begin{eqnarray}\label{e1}
 \frac{d {\bm \beta}}{d t}  
 &\simeq&   
   (1\!+\!3\epsilon^2\phi) \, \frac{e}{m}  {\bm \beta}\!\times\!{\bm B}  
,
\end{eqnarray}
where ${\bm \beta}\!=\!{\bm v}/c$, ${\bm v}=d{\bm x}/dt$ is the 3-velocity, $\epsilon=1/c$,
and $\phi\!=\!-GM/r$ is the gravitational potential.
Using this equation, they argue that the potential $\phi$ has a physical influence on the effective $g-2$.
In their argument, they seem to tacitly assume that time $t$ in (\ref{e1}) is the physical time. 
However, it is merely a coordinate time,
and not a physical time. When the equations are rewritten in terms of physical time and space coordinates, 
the dependence on $\phi$ disappears.

The essential reason why the dependence on $\phi$ disappears in physical coordinates can be understood 
in a simple way as follows. 
Let us write the metric tensor around Earth in the usual spherical coordinates as
\begin{equation}\label{e2}
ds^2=f(r)dt^2-f^{-1}(r)dr^2-r^2d\Omega^2 ,
\end{equation}
where 
\begin{equation}
f(r)=1-2GM/r=1+2\phi(r)
\end{equation}
and we take units $c=1$. For horizontal motion we have $dr=0$, so the metric above can be 
replaced by the effective metric
\begin{equation}\label{e3}
ds^2=f(r)dt^2-r^2d\Omega^2 ,
\end{equation}
where $r$ is treated as a constant. Introducing a new coordinate $t'$ defined by
\begin{equation}\label{e4}
 t'=\sqrt{f(r)}t ,
\end{equation}
the metric above can be written as
\begin{equation}\label{e5}
ds^2=dt'^2-r^2d\Omega^2 .
\end{equation}
Clearly, $t'$ is the physical time measured by a clock at the fixed radius $r$, while $t$ is merely a coordinate time. 
This means that (\ref{e5}) is the effective metric expressed in the physical coordinates, which does not depend on $\phi(r)$. 
In essence, this is why the horizontal motion, when expressed in terms of physical time and space coordinates,
cannot depend on the potential $\phi(r)$.

\section*{Acknowledgments}
 
The author is grateful to G. Duplan\v{c}i\'c, B. Klajn and B. Meli\'c for discussions.
This work was supported by the Ministry of Science of the Republic of Croatia
and by H2020 Twinning project No. 692194, ``RBI-T-WINNING''.

\end{document}